\documentclass[cits]{PoS}

\usepackage[authoryear,square]{natbib}
\bibpunct{(}{)}{;}{a}{}{,}

\title{Studies of Anomalous Microwave Emission (AME) with the SKA}

\ShortTitle{AME with the SKA}

\author{
\speaker{Clive Dickinson},$\!^{1}$ Y.\,Ali-Ha\"{i}moud,$\!^{2}$
R.\,J.\,Beswick,$\!^{1}$ S.\,Casassus,$\!^{3}$ K.\,Cleary,$\!^{4}$ B.\,T.\,Draine,$\!^5$ R.\,Genova-Santos,$\!^6$ K.\,Grainge,$\!^1$ T.\,C.\,Hoang,$\!^7$ A.\,Lazarian,$\!^8$ E.\,J.\,Murphy,$\!^9$ R.\,Paladini,$\!^9$ M.\,W.\,Peel,$\!^1$ Y.\,Perrott,$\!^{10}$ J.-A.\,Rubi\~{n}o-Martin,$\!^6$ A.\,Scaife,$\!^{11}$ C.\,T.\,Tibbs,$\!^{9}$ L.\,Verstraete,$\!^{12}$ M.\,Vidal,$\!^{1}$ R\,A.\,Watson,$\!^{1}$  N.\,Ysard$\!^{13}$\\ 
$^1$Jodrell Bank Centre for Astrophysics, The University of Manchester, UK\\
$^2$Institute for Advanced Study, Einstein Drive, Princeton, New Jersey 08540, USA \\
$^3$Departamento de Astronom\'{i}a, Universidad de Chile, Casilla 36-D, Santiago, Chile\\
$^4$Cahill Center for Astronomy \& Astrophysics, Caltech, Pasadena CA, 91125, USA\\
$^5$Princeton University Observatory, Princeton, NJ 08544-1001, USA\\
$^6$Instituto de Astrofi\'{s}ica de Canarias, 38200 La Laguna, Tenerife, Canary Islands, Spain\\
$^7$Institut f\"{u}r Theoretische Physik, Lehrstuhl IV: Weltraum- und Astrophysik, Bochum, Germany\\
$^8$Department of Astronomy, University of Wisconsin, Madison, WI 53706, USA\\
$^9$IPAC, Caltech, Pasadena CA, 91125, USA\\
$^{10}$Astrophysics Group, Cavendish Laboratory, University of Cambridge CB3 0HE, UK\\
$^{11}$School of Physics \& Astronomy, University of Southampton, Southampton, SO17 1BJ, UK \\
$^{12}$IAS, CNRS, Universit\'{e} Paris-Sud 11, Orsay, France \\
$^{13}$Department of Physics, PO Box 64, FI-00014 University of Helsinki, Finland, \\
E-mail: \email{Clive.Dickinson@manchester.ac.uk}
}

\abstract{
In this chapter, we will outline the scientific motivation for studying Anomalous Microwave Emission (AME) with the SKA. AME is thought to be due to electric dipole radiation from small spinning dust grains, although thermal fluctuations of magnetic dust grains may also contribute. Studies of this mysterious component would shed light on the emission mechanism, which then opens up a new window onto the interstellar medium (ISM). AME is emitted mostly in the frequency range $\sim 10$--100\,GHz, and thus the SKA has the potential of measuring the low frequency side of the AME spectrum, particularly in band 5. Science targets include dense molecular clouds in the Milky Way, as well as extragalactic sources. We also discuss the possibility of detecting rotational line emission from Poly-cyclic Aromatic Hydrocarbons (PAHs), which could be the main carriers of AME. Detecting PAH lines of a given spacing would allow for a definitive identification of specific PAH species.
}

\FullConference{
Advancing Astrophysics with the Square Kilometre Array\\
June 8-13, 2014\\
Giardini Naxos, Sicily, Italy}

\begin{document}

\section{Introduction}

Anomalous Microwave Emission (AME) is a mysterious component of microwave radiation emitted from the diffuse interstellar medium (ISM), which was first discovered in the late 1990s by sensitive CMB experiments operating at frequencies above 10\,GHz \citep{Leitch1997,Kogut1996}. Since then, it has been detected by numerous experiments operating in the frequency range $\sim 10$--$100$\,GHz, both in diffuse cirrus at high Galactic latitude \citep{deOliveira-Costa1999,deOliveira-Costa2004,Banday2003,Finkbeiner2004,Davies2006,Gold2011,Ghosh2012,Bennett2013} as well as from individual Galactic clouds \citep{Finkbeiner2002,Watson2005,Casassus2006,Dickinson2009a,Scaife2009,Scaife2010b,Dickinson2009a,Dickinson2010,Vidal2011,Planck2014_XV}. More recently it has been detected from individual star forming regions in a nearby galaxy \citep{Murphy2010,Scaife2010a}. The intensity is closely correlated with the thermal radiation at mid and far infrared (FIR) wavelengths associated with interstellar dust grains, but thermal dust emission cannot account for the observed microwave excess, which appears to peak at frequencies $\sim 20$--40\,GHz. Other traditional emission mechanisms (e.g., CMB, synchrotron, free-free) also cannot easily explain it. 

A number of models have been proposed to explain the AME. The most plausible is electric dipole radiation from ultra-small rapidly spinning dust grains (a.k.a. ``spinning dust''; see \S\,\ref{sec:spinningdust}). Low resolution {\it Planck} data and theoretical modelling of the Perseus and $\rho$~Ophiuchi molecular clouds show, beyond reasonable doubt, that the AME in these regions is indeed due to spinning dust \citep{Planck2011_XX}. However, a number of mysteries still remain. The classical picture of spinning dust grains does not appear to fit so well on small angular scales. Even though AME has been detected on arcmin scales \citep{Scaife2009,Tibbs2013}, the emission appears to be correlated with the strength of the radiation field rather than the column of small grains/PAHs along some sight-lines \citep{Tibbs2011,Tibbs2012}. Also, in some objects (such as compact HII regions), the data have failed to detect AME at all \citep{Scaife2008}. The AME could also be due to another mechanism, such as thermal fluctuations in magnetic grains (a.k.a. ``magnetic dust''; see \S\,\ref{sec:magneticdust}). Detailed observations of AME will help us understand this mysterious component and, in principle, open up a new window on the ISM, both in the Milky Way and in external galaxies.

In this chapter, we begin by reviewing the two main models for AME (\S\,\ref{sec:models}) and what the predictions are for their radio emission. We also mention the possibility of detecting rotational PAH lines in AME regions. We then discuss potential targets (\S\,\ref{sec:targets}) and the instrumental requirements (\S\,\ref{sec:requirements}) that would allow the SKA to observe AME and what we could learn from them. We conclude and summarise the outlook for AME observations with the SKA (\S\,\ref{sec:conclusions}).

%%%%%%%%%%%%%%%%%%%%%%%%%%%%%%%%%%%%%%%%%%%%%%%%%%%%%%%%%%%%%%%%%%%%%%%%%
%%%%%%%%%%%%%%%%%%%%%%%%%%%%%%%%%%%%%%%%%%%%%%%%%%%%%%%%%%%%%%%%%%%%%%%%%

\section{Models for the AME - Brief Overview and Predictions}
\label{sec:models}

AME was given its name \citep{Leitch1997} because it could not easily be explained in terms of conventional physical emission mechanisms - synchrotron, free-free, thermal dust, or CMB emissions. A number of possibilities have been explored, including flat spectrum synchrotron radiation \citep{Bennett2003b,Peel2012} and hot ($T_e \gtrsim 10^6$\,K) free-free emission \citep{Leitch1997}. There are currently two plausible explanations for the AME - electric dipole radiation from small rapidly spinning dust grains and thermal fluctuations in magnetic dust grains. Although spinning dust is the most widely accepted explanation there is still a possibility that some, or all, of the AME is from magnetic dust. Here we briefly review the theory and predictions. Finally, we also mention the possibility of detecting rotational line emission from PAHs, which has recently been proposed\footnote{Note that the line emission is in addition to continuum emission from spinning dust grains.} \citep{Ali-Hamoud2014}.

\subsection{Spinning dust}
\label{sec:spinningdust}

Electric dipole radiation from small spinning dust grains has been predicted for a long time \citep{Erickson1957}. However, it had not received much attention until the late 1990s when several experiments detected excess emission at high frequencies (10--60\,GHz). The first detailed theory and predictions were made by \cite{Draine1998b}, with subsequent minor improvements over recent  years \citep{Ali-Hamoud2009,Ysard2010a,Hoang2010,Hoang2011}.

The idea is simple - rotating electric dipoles radiate. Interstellar dust grains will naturally possess permanent electric dipoles, $\mu$, due to an uneven distribution of charges, impurities (such as nitrogen in PAHs), or an asymmetric structure; even spherically symmetric grains will have, on average, a residual dipole moment. Also, interstellar dust grains are known to spin at high rates due to a number of excitation mechanisms (photons, collisions, formation of molecules etc). Classically, the power radiated by a dipole rotating at angular velocity $\omega$, with electric dipole component $\mu_{\perp}$ perpendicular to $\omega$ is given by

\begin{equation}
P = \frac{2}{3} \frac{\mu_{\perp}^2 \omega^4}{c^3} ~.
\end{equation}
This power is emitted at a frequency $\nu=\omega / 2\pi$. To calculate the detailed frequency spectrum of spinning dust, one needs to integrate the emission over a distribution of grain sizes, electric dipole moments, and angular velocities. Specifically, the emissivity per H atom is \citep{Ali-Hamoud2009}:

\begin{equation}
\frac{j_{\nu}}{n_{\rm H}} = \frac{1}{4\pi} \int_{a_{\rm min}}^{a_{\rm max}} da \frac{1}{n_{\rm H}} \frac{dn_{\rm gr}}{da} 4\pi\omega^2 f_{\rm a}(\omega) 2\pi \frac{2}{3} \frac{\mu^2_{a\perp} \omega^4}{c^3},
\end{equation}
where $\omega=2\pi\nu$, $dn_{\rm gr}/da$ is the grain size distribution, $f_{\rm a}(\omega)$ is the rotational velocity distribution.

\begin{figure}[!h]
\centering
\includegraphics[width=.45\textwidth]{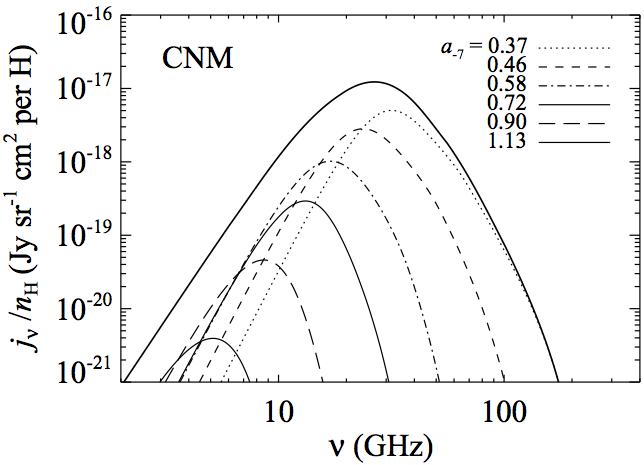}
\caption{Theoretical spinning dust emissivity for a CNM environment (thick solid line). Contributions from grains of various sizes are shown [$a_7 \equiv a/(10^{-7}$\,cm]. This spectrum is typical of spinning dust models, which show a highly peaked spectrum, with a peak frequency $\sim 30$\,GHz, but is very sensitive to the properties of the grain and environment. Reproduced from \cite{Ali-Hamoud2009}.}
\label{fig:spdust_spec}
\end{figure}

It is easy to see that for grains to emit appreciable power at radio frequencies, the grains must spin at a very high rate - at GHz frequencies. Larger grains, due to their larger moments-of-inertia, emit at lower frequencies and with a lower emissivity. At frequencies $<5$\,GHz, spinning dust is expected to be very weak, and typically swamped by the usual free-free and synchrotron components.  The high frequency cut-off is determined by the very smallest grains (VSGs) and PAHs, which are the most emissive and dominate the spectrum, resulting in a peaked spectrum. Fig.~\ref{fig:spdust_spec} shows a theoretically predicted spectrum for specific conditions typical of the Cold Neutral Medium (CNM). It can be seen that most of the emission comes from the very smallest grains.

\begin{figure}[!h]
\centering
\includegraphics[width=.40\textwidth]{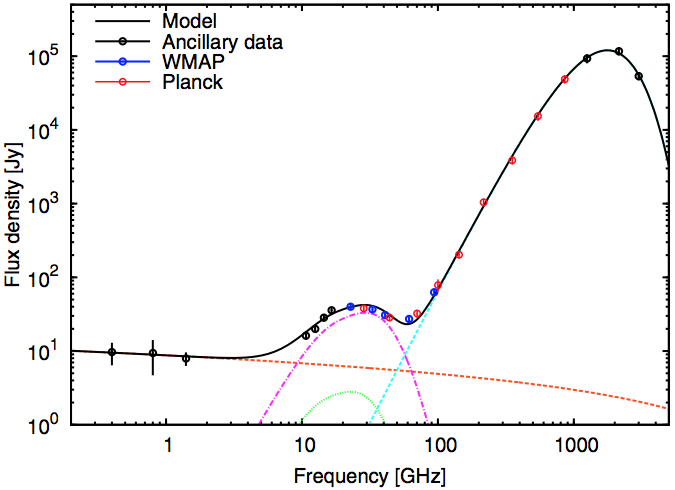}
\includegraphics[width=.40\textwidth]{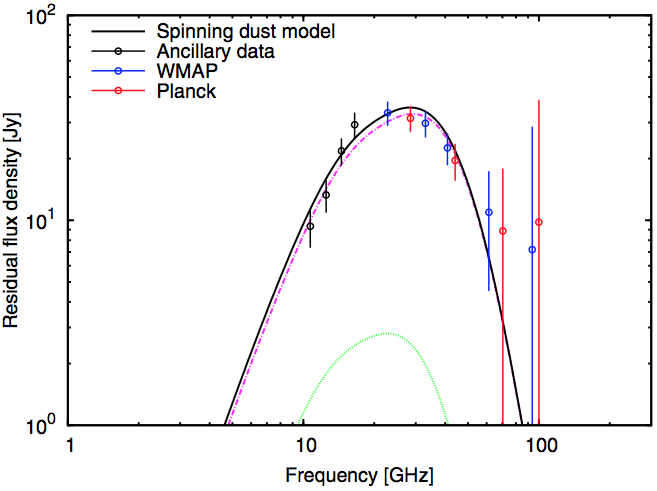}
\caption{The spectrum of G160.26--18.62 in the Perseus molecular cloud ({\it left}) and the residual spectrum showing the spinning dust component ({\it right}). The spectrum is fitted by components of free-free (orange dashed line), CMB (not visible), thermal dust (dashed cyan line) and spinning dust (green dotted and magenta dot-dashed lines for the atomic and molecular phases, respectively), which peaks at $\approx 30$\,GHz. The spectrum is a remarkably good fit to the data with parameters that are physically motivated. Reproduced from \cite{Planck2011_XX}.}
\label{fig:perseus_spec}
\end{figure}

Although there is still significant debate about the origin of the diffuse high latitude AME observed in CMB experiments \citep{Bennett2003b}, there are a few examples of Galactic molecular clouds that are now widely accepted to emit significant amounts of spinning dust at frequencies $\sim 30$\,GHz. The best two examples are the Perseus and $\rho$\,Ophiuchi molecular clouds; Fig.~\ref{fig:perseus_spec} shows the SED of G160.26--18.62 in the Perseus molecular cloud and the spinning dust residuals including {\it Planck} data \citep{Planck2011_XX}. The spinning dust is modelled by 2 components representing the atomic and molecular phases of the ISM. The analysis not only showed that the spectrum can be well-fitted by spinning dust but also that the physical parameters that define the model are consistent with expectations i.e. the model is physically motivated, or alternatively, that we expect to see spinning dust at this  level in this region. In fact, a number of new observations are now suggesting that a significant fraction (tens of percent) of the 30\,GHz emission in our Galaxy is due to AME \citep{Planck2014_XXIII}. This could have significant consequences for astronomical observations of our Galaxy and extragalactic sources\footnote{The study of AME may be of relevance to studying star formation in galaxies via their free-free emission \citep{Beswick2014,Murphy2014}, since a separation between the synchrotron/free-free/AME components is required above $\sim 10$\,GHz.}

Spinning dust emission is largely unpolarized because the smallest grains that dominate the power do not align efficiently with the magnetic field. At 30\,GHz the polarized fraction is expected to be $<1\,\%$ \citep{Lazarian2000,Hoang2013}, consistent with current upper limits \citep{Dickinson2011,Lopez-Caraballo2011,Rubino-Martin2012a}. At lower frequencies, where larger grains contribute, the polarization can be significant (few \,\%) but will be difficult to measure due to confusion with synchrotron/free-free emission. Nevertheless, measuring polarization would provide constraints on paramagnetic alignment and resonance relaxation models of small grain alignment \citep{Lazarian2000} and, perhaps more importantly, may allow us to distinguish between spinning dust (low polarization) and magnetic dust (\S\,\ref{sec:magneticdust}) mechanisms (high polarization).

\subsection{Magnetic dust}
\label{sec:magneticdust}

An alternative explanation for AME was proposed by \cite{Draine1999} in terms of thermal fluctuations of the magnetization within individual interstellar grains. Such thermal fluctuations result in magnetic dipole radiation. The predictions of magnetic dust are based on the magnetic properties of dust using the Gilbert equation \citep{Draine2012} and are quite diverse. Although ordinary paramagnetic grains cannot account for the bulk of the AME, stronger magnetic dipole radiation will result if a fraction of the grain material is ferromagnetic by the inclusion of iron (Fe). Since $\sim 90\,\%$ of the Fe is missing from the gas phase, it must be locked up in solid grains; very little is known about the nature of the Fe-containing material. Again, the predictions depend critically on the form of Fe e.g. as free flyers, inclusions on grains, metallic Fe, magnetite, maghemite etc. \citep{Draine2013}.

\begin{figure}[!h]
\centering
\includegraphics[width=.40\textwidth]{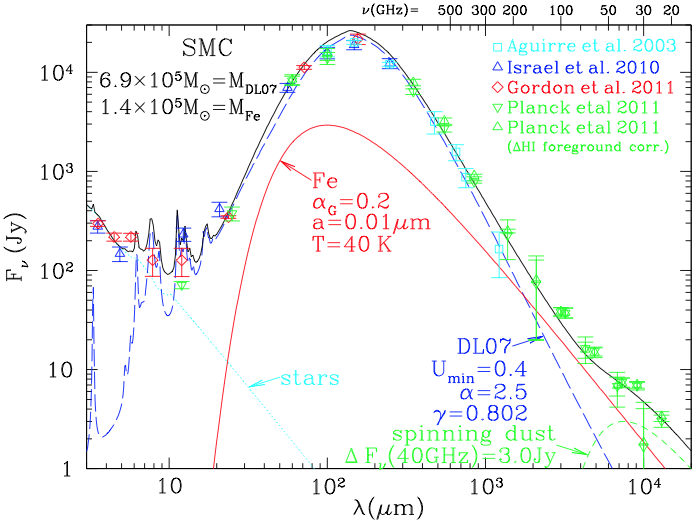}
\includegraphics[width=.40\textwidth]{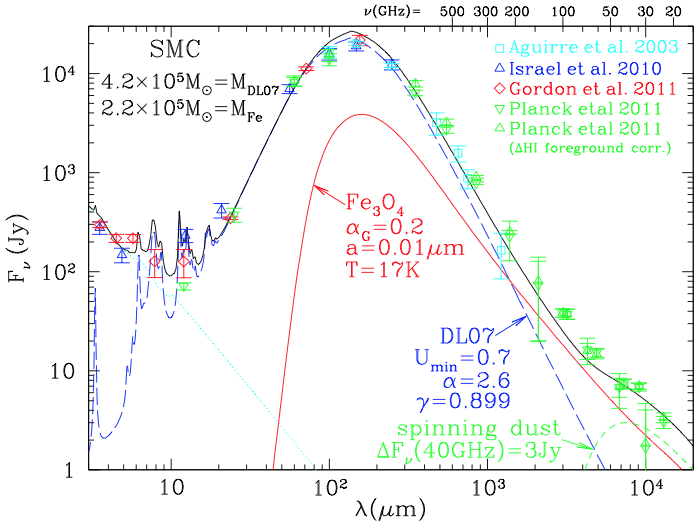}
\caption{The spectrum of the SMC, fitted with two example models including magnetic dust. The excess emission relative to a simple thermal dust model is observed at frequencies $\sim 50$--200\,GHz. In these fits, a combination of spinning dust  (dashed green line) and magnetic dust (red line) can explain most of the excess. The plots show two different example models from metallic Fe at $T=40$\,K ({\it left}) and magnetite at $T=17$\,K ({\it right}). Reproduced from \cite{Draine2012}.}
\label{fig:smc_spec}
\end{figure}

Magnetic dust has not been definitively detected as of yet. However, a number of hints have been indicated. A study of the sub-mm spectrum of the Small Magellanic Cloud (SMC) has shown that the sub-mm excess can be readily accounted for by magnetic dust \citep{Draine2012}. Fig.~\ref{fig:smc_spec} shows the spectrum of the SMC, fitted with two models including a contribution from magnetic dust. It is worth noting that the magnetic dust does not depend sensitively on the size of grains (unlike for spinning dust) and thus it could emit from anywhere there is dust.

The polarization of magnetic dust is similarly complicated and depends on the form of the magnetic nanoparticles. Most models predict a very high ($\sim 35\,\%$) polarization in the sub-mm, decreasing towards microwave/radio wavelengths. Indeed, a recent analysis of {\it Planck} polarization data shows a fall in the polarized fraction from 353\,GHz to 100\,GHz, which may indicate a component of magnetic dust \citep{Planck2014_XXII}. This polarization signature is the most clear way to distinguish between magnetic dust and other emission mechanisms such as spinning and thermal dust. Note that the polarization undergoes a reversal at frequencies $\sim 3$--20\,GHz, depending on grain geometry and composition i.e. the emission will go from being aligned with the magnetic field to being perpendicular to it \citep{Draine2012}. This would be a smoking gun indication of this process, although confusion from the polarized synchrotron radiation will likely make this difficult.

\subsection{Rotational spectroscopy of interstellar PAHs}
\label{sec:lines}

Carbonaceous dust grains are thought to extend down to the molecular regime at the small size end, where they take the form of Poly-cyclic Aromatic Hydrocarbons (PAHs) \citep{Tielens2008}. PAHs are widely recognised as the carriers of the near-infrared aromatic features, are abundant in the interstellar medium, and are important actors in its chemical and thermal balance. They are also presumably the emitters of spinning dust radiation. Yet, no specific interstellar PAH has been identified to date, as the near-infrared features are not molecule-specific. Detecting and identifying interstellar PAHs would close any remaining doubt about the nature of the carriers of near-infrared features \citep{Kwok2013} and could lead to a more quantitative understanding of their formation mechanisms.

\begin{figure}[!h]
\centering
\includegraphics[width=0.7\textwidth]{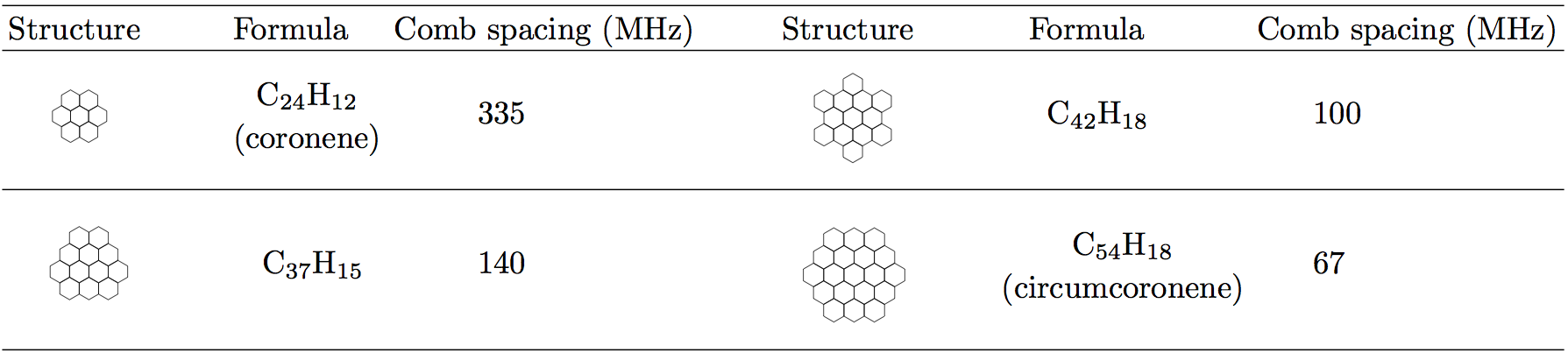}
\caption{Approximate line spacing for representative symmetric PAHs, computed assuming a perfectly hexagonal carbon skeleton with a C-C bond length of $1.4$\,\AA~and a C-H bond length of $1.1$\,\AA~for the peripheral hydrogen atoms.}
\label{fig:pahs}
\end{figure}

It was recently suggested that rotational spectroscopy combined with matched filtering techniques could be used to identify interstellar PAHs \citep{Ali-Hamoud2014}. General PAHs are large asymmetric molecules and have a very complex rotational spectrum, unusable for identification purposes. However, quasi-symmetric PAHs (such as nitrogen-substituted coronene or circum-coronone; \citealt{Hudgins2005}) have very regular rotational spectra, taking the shape of "combs" of evenly spaced lines with width of order of a MHz\footnote{Each "tooth" of the comb spectrum is really a stack of lines spread across $\sim 1$\,MHz. Observed with a $\sim$\,MHz resolution these stacks would appear as lines.}. The line separation ranges from a few tens of MHz to a few hundreds of MHz and is highly sensitive to the size of the emitting species; see the table in Fig.~\ref{fig:pahs} for the approximate line spacing of a few representative species. 

The expected strength of the lines is highly uncertain, as it depends of the fraction of PAHs that are quasi-symmetrical. If of order one percent of PAH molecules are in the form of coronene, for example, we expect a line strength of order a mJy (near the peak of AME, at $\sim 10$--30\,GHz) for a molecular cloud observed with ~arcmin resolution. The highly regular shape of the spectrum allows for a more efficient search using matched filtering techniques.

%%%%%%%%%%%%%%%%%%%%%%%%%%%%%%%%%%%%%%%%%%%%%%%%%%%%%%%%%%%%%%%%%%%%%%%%%
%%%%%%%%%%%%%%%%%%%%%%%%%%%%%%%%%%%%%%%%%%%%%%%%%%%%%%%%%%%%%%%%%%%%%%%%%

\section{Observations of AME with the SKA}

\subsection{Potential targets}
\label{sec:targets}

AME from spinning and magnetic dust radiation is expected, at some level, from any object containing small dust grains. Therefore, there are potentially a very wide range of objects that could be studied, including molecular clouds, photodissociation regions (PDRs), dense cores, circumstellar disks etc. However, some consideration is needed for the contribution of synchrotron/free-free since the SKA will operate at $<20$\,GHz where these are significant for many objects. Furthermore, many of these sources are very diffuse with the emission being distributed on scales of arcmin and larger. This will cause difficulties in imaging the large-scale emission (resolving out flux, deconvolution etc.). Nevertheless, ISM structures invariably emit over a wide range of scales and we would need to focus on regions that have a significant column of dust, using ancillary IR/sub-mm data. 

One of the best targets would be dense pre-stellar cores. These collapsing dust clouds have typical sizes of $\sim 1$\,arcmin or less, which is well matched to the shorter SKA baselines. Although a large fraction of these are found within large molecular cloud complexes with strong synchrotron/free-free emission, some are less contaminated and have clean lines-of-sight. There is an interesting possibility of using the profile of spinning dust emission as a probe of these clouds. \cite{Ysard2011} have shown that this could in principle be used to study the grain properties and environment within the cloud. In particular, they proposed a scheme to estimate the abundances of the major ions (H{\sc ii}, C{\sc ii}), which would be complementary to direct observations. The emission may also be relatively strong at the centre of dense clouds where the mid-IR PAH emission is weak and could provide a new way to trace grain growth from diffuse to dense media. 

Circumstellar disks would also be a good target for AME. Near-IR studies (and more recently results from the Atacama Large Millimeter Array; ALMA) of disks around Herbig Ae/Be stars have demonstrated that a substantial amount of dust mass can be locked up in nanometre-sized particles. Modelling has shown that spinning dust could contribute of order mJy at $\sim 20$\,GHz \citep{Rafikov2006} and hence at levels comparable to the Rayleigh-Jeans tail of the sub-mm continuum. SKA would provide high resolution imaging of disks and the separation from synchrotron/free-free could be achieved by observing at lower frequencies. Detection of the spinning dust emission would provide important evidence for the existence, properties, and origin of the population of small dust particles in proto-planetary disks, with possible ramifications for planet formation. VSGs are key to the thermal balance of proto-planetary disks, and their detection at radio frequencies would allow a new probe of their distribution,  unhindered by the central circumstellar glare that affects the IR observations.

Another obvious target for the high resolution capabilities of SKA, is external galaxies (see also \citealt{Murphy2014}). Strangely, AME has only been detected in one external galaxy, NGC6946 \citep{Murphy2010,Scaife2010a} (and possibly SMC/LMC). In fact, it was only significantly detected in two out of ten star forming regions within the spiral arms of NGC6946. It is possible that AME may only be strongly emitted in a small volume of an entire galaxy, and thus the integrated contribution of AME could be small in general. On the other hand, there have been few detailed measurements at frequencies above a few GHz, thus we may have simply missed it. Another interesting possibility is that high redshift galaxies ($z > 2$), would emit AME at observed frequencies below $\sim 10$\,GHz, which may be more suited to the SKA. The higher star formation rates, and therefore increased dust production at these redshifts, may result in a relatively larger fraction of AME.

For the detection of rotational PAH lines, we would target regions that are known to have column density of PAHs (via IR maps e.g. {\it Spitzer} 8\,$\mu$m) but little ionized gas to reduce contamination from radio recombination and other ionic lines. AME-dominated targets would be obvious targets.

\subsection{Instrument requirements}
\label{sec:requirements}

AME, whether from spinning or magnetic dust, emits at relatively high radio (microwave) frequencies - at $\sim 10$\,GHz and above. The ideal instrument would cover frequencies of a few GHz up to $\sim 300$\,GHz. However, there is a dearth of observations in the range 10--30\,GHz. The SKA would be complementary to sub-mm data (e.g. ALMA and {\it Planck}) and would provide the crucial low frequency data required to separate synchrotron/free-free/AME. 

In SKA1-MID, the highest frequency corresponds to band 5 covering 4.6--13.8\,GHz and with a bandwidth of 2.5\,GHz in each polarisation \citep{Dewdney2013}. For spinning dust, the focus would be on intensity rather than polarization, while for magnetic dust the polarization would be the main interest. Polarization observations either side of the predicted $\sim 15$\,GHz reversal (e.g. combined with ALMA band 1 30\,GHz observations or Very Large Array) would provide a smoking gun detection of magnetic nanoparticles. The full SKA is expected to include the highest frequency band 5 and may go up to 24\,GHz, or possibly even a little higher. Clearly this would open up the SKA to studies of AME, both spinning dust and magnetic dust. Assuming a build-out from the central core of SKA1-MID, bright AME targets offer the opportunity for early science with $<50\,\%$ capability of the full SKA1-MID.

The angular scale coverage is more difficult to define since there is a wide range of potential targets such as prestellar cores, diffuse molecular clouds and extragalactic sources (\S\,\ref{sec:targets}). An important consideration is the surface brightness sensitivity since the majority of AME detections have been from diffuse regions rather than from compact sources. We therefore require good $u,v$ coverage with many short baselines. For continuum, SKA1-MID is likely to provide sufficient sensitivity for many of our targets, which would be chosen carefully to match the $u,v$ coverage that was available. Utilising the lower bands of SKA (band 3 [1.6--3.1\,GHz] and band 4 [2.8--5.2\,GHz]), using matched baselines to give equivalent $u,v$-coverage, could be used to differentiate AME from synchrotron/free-free emissions.

To appreciate the sensitivity and power of the SKA, we make an order-of-magnitude estimate of the integration time required for SKA-MID to detect AME in one of the well-known diffuse molecular clouds. We consider the filamentary structure observed in the Perseus molecular cloud by the Arc Minute Imager (AMI) at 16\,GHz and synthesized beam of $\approx 2$\,arcmin. AME structure was detected on the scale of the beam at a level of $\approx 6$\,mJy/beam \citep{Tibbs2013}. If we conservatively assume that the emission is completely smooth (does not have structure on scales smaller than the beam), we can calculate the lower limit to the flux density observed with the 1\,km core of SKA-MID, which corresponds to an angular resolution of $\sim 3$\,arcmin, of $\sim 4\,\mu$Jy/beam. For early observations with $50\,\%$ early phase 1 SKA, the system equivalent flux density is 3.4\,Jy (1.7\,Jy for the full array; \citealt{Dewdney2013}). For a bandwidth of 2.5\,GHz, an integration time of 2300\,secs ($\sim 1$\,hour) is required to achieve a $5\,\sigma$ detection. It is therefore clear that even for diffuse ISM structures, the SKA can potentially be very powerful.

For the rotational lines of PAHs, we require a frequency resolution of $\sim 1$\,MHz or better, which would be easily achievable with 16384 channels over a 5\,GHz bandwidth centred at $\approx 20$\,GHz. The required sensitivity of $\sim \mu$Jy/beam is readily feasible as long as the ISM structures were matched to the SKA baselines, while the stacking analysis can be made to be robust against bandpass and other calibration errors \citep{Ali-Hamoud2014}.

%%%%%%%%%%%%%%%%%%%%%%%%%%%%%%%%%%%%%%%%%%%%%%%%%%%%%%%%%%%%%%%%%%%%%%%%%
%%%%%%%%%%%%%%%%%%%%%%%%%%%%%%%%%%%%%%%%%%%%%%%%%%%%%%%%%%%%%%%%%%%%%%%%%

\section{Conclusions and outlook}
\label{sec:conclusions}

SKA, operating at frequencies above $\sim 10$\,GHz would provide high resolution (arcsec and better) and high fidelity wide-field imaging of Galactic clouds and similar regions in nearby galaxies. Precise continuum measurements of the frequency spectrum of AME would provide definitive confirmation of the spinning dust model across a wide range of scales. Sensitive and high resolution observations of molecular dust clouds, away from hot ionized gas (which is invariably associated with star forming clouds), would give a clear picture of AME. Combined with higher frequency data (e.g. ALMA band 1 at 30\,GHz) we can establish the full spinning dust SED for a variety of targets. 

Once spinning dust is confirmed, we can then use this unique tracer of small dust grains as a new probe of the ISM. The spinning dust spectrum and intensity is very sensitive to a number of parameters associated with the ISM, that are otherwise difficult to constrain. Combined with theoretical modelling, it would allow the dust grain size distribution to be measured as well as the density and interstellar radiation field (and possibly other parameters such as the electric dipole moment), which affect the rotational excitation of interstellar grains. Models already exist for cold prestellar cores, which predict specific signatures of spinning dust that can be tested and unique information on the physics of star-forming regions can be gained. Complementary high resolution FIR/sub-mm data (e.g. {\it Spitzer/Herschel}, ALMA) would provide a complete picture of the ISM, which could be studied on a wide range of scales and environments.

Finally, we outline the possibility of detecting comb-like line emission from PAHs, which would provide definitive identification of specific, free-floating interstellar PAHs.

CD acknowledges support from an ERC Starting (Consolidator) Grant (no~307209).

%%%%%%%%%%%%%%%%%%%%%%%%%%%%%%%%%%%%%%%%%%%%%%%%%%%%%%%%%%%%%%%%%%%%%%%%%
%%%%%%%%%%%%%%%%%%%%%%%%%%%%%%%%%%%%%%%%%%%%%%%%%%%%%%%%%%%%%%%%%%%%%%%%%

\setlength{\bibsep}{0.0pt}

\bibliography{clive_refs}{}

\begin{thebibliography}{}
\expandafter\ifx\csname natexlab\endcsname\relax\def\natexlab#1{#1}\fi

\bibitem[{{Ali-Ha{\"i}moud}(2014)}]{Ali-Hamoud2014}
{Ali-Ha{\"i}moud}, Y. 2014, \mnras, 437, 2728

\bibitem[{{Ali-Ha{\"\i}moud} {et~al.}(2009){Ali-Ha{\"\i}moud}, {Hirata}, \&
  {Dickinson}}]{Ali-Hamoud2009}
{Ali-Ha{\"\i}moud}, Y., {Hirata}, C.~M., \& {Dickinson}, C. 2009, \mnras, 395,
  1055

\bibitem[{{Banday} {et~al.}(2003){Banday}, {Dickinson}, {Davies}, {Davis}, \&
  {G{\'o}rski}}]{Banday2003}
{Banday}, A.~J., {Dickinson}, C., {Davies}, R.~D., {Davis}, R.~J., \&
  {G{\'o}rski}, K.~M. 2003, \mnras, 345, 897

\bibitem[{{Bennett} {et~al.}(2003){Bennett}, {Hill}, {Hinshaw}, {Nolta},
  {Odegard}, {Page}, {Spergel}, {Weiland}, {Wright}, {Halpern}, {Jarosik},
  {Kogut}, {Limon}, {Meyer}, {Tucker}, \& {Wollack}}]{Bennett2003b}
{Bennett}, C.~L., {Hill}, R.~S., {Hinshaw}, G., {et~al.} 2003, \apjs, 148, 97

\bibitem[{{Bennett} {et~al.}(2012){Bennett}, {Larson}, {Weiland}, {Jarosik},
  {Hinshaw}, {Odegard}, {Smith}, {Hill}, {Gold}, {Halpern}, {Komatsu}, {Nolta},
  {Page}, {Spergel}, {Wollack}, {Dunkley}, {Kogut}, {Limon}, {Meyer}, {Tucker},
  \& {Wright}}]{Bennett2013}
{Bennett}, C.~L., {Larson}, D., {Weiland}, J.~L., {et~al.} 2012, accepted in
  ApJSS, (ArXiv:1212.5225), arXiv:1212.5225

\bibitem[{{Beswick}(2014)}]{Beswick2014}
{Beswick}, R.~J. 2014, in Advancing Astrophysics with the Square Kilometre
  Array, PoS(AASKA14)070

\bibitem[{{Casassus} {et~al.}(2006){Casassus}, {Cabrera}, {F{\"o}rster},
  {Pearson}, {Readhead}, \& {Dickinson}}]{Casassus2006}
{Casassus}, S., {Cabrera}, G.~F., {F{\"o}rster}, F., {et~al.} 2006, \apj, 639,
  951

\bibitem[{{Davies} {et~al.}(2006){Davies}, {Dickinson}, {Banday}, {Jaffe},
  {G{\'o}rski}, \& {Davis}}]{Davies2006}
{Davies}, R.~D., {Dickinson}, C., {Banday}, A.~J., {et~al.} 2006, \mnras, 370,
  1125

\bibitem[{{de Oliveira-Costa} {et~al.}(2004){de Oliveira-Costa}, {Tegmark},
  {Davies}, {Guti{\'e}rrez}, {Lasenby}, {Rebolo}, \&
  {Watson}}]{deOliveira-Costa2004}
{de Oliveira-Costa}, A., {Tegmark}, M., {Davies}, R.~D., {et~al.} 2004, \apjl,
  606, L89

\bibitem[{{de Oliveira-Costa} {et~al.}(1999){de Oliveira-Costa}, {Tegmark},
  {Gutierrez}, {Jones}, {Davies}, {Lasenby}, {Rebolo}, \&
  {Watson}}]{deOliveira-Costa1999}
{de Oliveira-Costa}, A., {Tegmark}, M., {Gutierrez}, C.~M., {et~al.} 1999,
  \apjl, 527, L9

\bibitem[{{Dewdney} {et~al.}(2013){Dewdney}, {Turner}, {Millenaar}, {McCool},
  {Lazio}, \& {Cornwell}}]{Dewdney2013}
{Dewdney}, P.~E., {Turner}, W., {Millenaar}, R., {et~al.} 2013, SKA1 System
  Baseline Design, Tech. Rep. SKA-TEL-SKO-DD-001 Revision 1, SKA Organisation

\bibitem[{{Dickinson} {et~al.}(2011){Dickinson}, {Peel}, \&
  {Vidal}}]{Dickinson2011}
{Dickinson}, C., {Peel}, M., \& {Vidal}, M. 2011, \mnras, 418, L35

\bibitem[{{Dickinson} {et~al.}(2009){Dickinson}, {Davies}, {Allison}, {Bond},
  {Casassus}, {Cleary}, {Davis}, {Jones}, {Mason}, {Myers}, {Pearson},
  {Readhead}, {Sievers}, {Taylor}, {Todorovi{\'c}}, {White}, \&
  {Wilkinson}}]{Dickinson2009a}
{Dickinson}, C., {Davies}, R.~D., {Allison}, J.~R., {et~al.} 2009, \apj, 690,
  1585

\bibitem[{{Dickinson} {et~al.}(2010){Dickinson}, {Casassus}, {Davies},
  {Allison}, {Bustos}, {Cleary}, {Davis}, {Jones}, {Pearson}, {Readhead},
  {Reeves}, {Taylor}, {Tibbs}, \& {Watson}}]{Dickinson2010}
{Dickinson}, C., {Casassus}, S., {Davies}, R.~D., {et~al.} 2010, \mnras, 407,
  2223

\bibitem[{{Draine} \& {Hensley}(2012)}]{Draine2012}
{Draine}, B.~T., \& {Hensley}, B. 2012, \apj, 757, 103

\bibitem[{{Draine} \& {Hensley}(2013)}]{Draine2013}
---. 2013, \apj, 765, 159

\bibitem[{{Draine} \& {Lazarian}(1998)}]{Draine1998b}
{Draine}, B.~T., \& {Lazarian}, A. 1998, \apj, 508, 157

\bibitem[{{Draine} \& {Lazarian}(1999)}]{Draine1999}
---. 1999, \apj, 512, 740

\bibitem[{{Erickson}(1957)}]{Erickson1957}
{Erickson}, W.~C. 1957, \apj, 126, 480

\bibitem[{{Finkbeiner}(2004)}]{Finkbeiner2004}
{Finkbeiner}, D.~P. 2004, \apj, 614, 186

\bibitem[{{Finkbeiner} {et~al.}(2002){Finkbeiner}, {Schlegel}, {Frank}, \&
  {Heiles}}]{Finkbeiner2002}
{Finkbeiner}, D.~P., {Schlegel}, D.~J., {Frank}, C., \& {Heiles}, C. 2002,
  \apj, 566, 898

\bibitem[{{Ghosh} {et~al.}(2012){Ghosh}, {Banday}, {Jaffe}, {Dickinson},
  {Davies}, {Davis}, \& {Gorski}}]{Ghosh2012}
{Ghosh}, T., {Banday}, A.~J., {Jaffe}, T., {et~al.} 2012, \mnras, 422, 3617

\bibitem[{{Gold} {et~al.}(2011){Gold}, {Odegard}, {Weiland}, {Hill}, {Kogut},
  {Bennett}, {Hinshaw}, {Chen}, {Dunkley}, {Halpern}, {Jarosik}, {Komatsu},
  {Larson}, {Limon}, {Meyer}, {Nolta}, {Page}, {Smith}, {Spergel}, {Tucker},
  {Wollack}, \& {Wright}}]{Gold2011}
{Gold}, B., {Odegard}, N., {Weiland}, J.~L., {et~al.} 2011, \apjs, 192, 15

\bibitem[{{Hoang} {et~al.}(2010){Hoang}, {Draine}, \& {Lazarian}}]{Hoang2010}
{Hoang}, T., {Draine}, B.~T., \& {Lazarian}, A. 2010, \apj, 715, 1462

\bibitem[{{Hoang} {et~al.}(2011){Hoang}, {Lazarian}, \& {Draine}}]{Hoang2011}
{Hoang}, T., {Lazarian}, A., \& {Draine}, B.~T. 2011, \apj, 741, 87

\bibitem[{{Hoang} {et~al.}(2013){Hoang}, {Lazarian}, \& {Martin}}]{Hoang2013}
{Hoang}, T., {Lazarian}, A., \& {Martin}, P.~G. 2013, \apj, 779, 152

\bibitem[{{Hudgins} {et~al.}(2005){Hudgins}, {Bauschlicher}, \&
  {Allamandola}}]{Hudgins2005}
{Hudgins}, D.~M., {Bauschlicher}, Jr., C.~W., \& {Allamandola}, L.~J. 2005,
  \apj, 632, 316

\bibitem[{{Kogut} {et~al.}(1996){Kogut}, {Banday}, {Bennett}, {Gorski},
  {Hinshaw}, {Smoot}, \& {Wright}}]{Kogut1996}
{Kogut}, A., {Banday}, A.~J., {Bennett}, C.~L., {et~al.} 1996, \apjl, 464, L5

\bibitem[{{Kwok} \& {Zhang}(2013)}]{Kwok2013}
{Kwok}, S., \& {Zhang}, Y. 2013, \apj, 771, 5

\bibitem[{{Lazarian} \& {Draine}(2000)}]{Lazarian2000}
{Lazarian}, A., \& {Draine}, B.~T. 2000, \apjl, 536, L15

\bibitem[{{Leitch} {et~al.}(1997){Leitch}, {Readhead}, {Pearson}, \&
  {Myers}}]{Leitch1997}
{Leitch}, E.~M., {Readhead}, A.~C.~S., {Pearson}, T.~J., \& {Myers}, S.~T.
  1997, \apjl, 486, L23

\bibitem[{{L{\'o}pez-Caraballo} {et~al.}(2011){L{\'o}pez-Caraballo},
  {Rubi{\~n}o-Mart{\'{\i}}n}, {Rebolo}, \&
  {G{\'e}nova-Santos}}]{Lopez-Caraballo2011}
{L{\'o}pez-Caraballo}, C.~H., {Rubi{\~n}o-Mart{\'{\i}}n}, J.~A., {Rebolo}, R.,
  \& {G{\'e}nova-Santos}, R. 2011, \apj, 729, 25

\bibitem[{{Murphy}(2014)}]{Murphy2014}
{Murphy}, E.~J. 2014, in Advancing Astrophysics with the Square Kilometre
  Array, PoS(AASKA14)085

\bibitem[{{Murphy} {et~al.}(2010){Murphy}, {Helou}, {Condon}, {Schinnerer},
  {Turner}, {Beck}, {Mason}, {Chary}, \& {Armus}}]{Murphy2010}
{Murphy}, E.~J., {Helou}, G., {Condon}, J.~J., {et~al.} 2010, \apjl, 709, L108

\bibitem[{{Peel} {et~al.}(2012){Peel}, {Dickinson}, {Davies}, {Banday},
  {Jaffe}, \& {Jonas}}]{Peel2012}
{Peel}, M.~W., {Dickinson}, C., {Davies}, R.~D., {et~al.} 2012, \mnras, 424,
  2676

\bibitem[{{Planck Collaboration} {et~al.}(2011){Planck Collaboration}, {Ade},
  {Aghanim}, {Arnaud}, {Ashdown}, {Aumont}, {Baccigalupi}, {Balbi}, {Banday},
  {Barreiro}, \& et~al.}]{Planck2011_XX}
{Planck Collaboration}, {Ade}, P.~A.~R., {Aghanim}, N., {et~al.} 2011, \aap,
  536, A20

\bibitem[{{Planck Collaboration} {et~al.}(2013){Planck Collaboration}, {Ade},
  {Aghanim}, {Alves}, {Arnaud}, {Atrio-Barandela}, {Aumont}, {Baccigalupi},
  {Banday}, {Barreiro}, {Battaner}, {Benabed}, {Benoit-L{\'e}vy}, {Bernard},
  {Bersanelli}, {Bielewicz}, {Bobin}, {Bonaldi}, {Bond}, {Borrill}, {Bouchet},
  {Boulanger}, {Burigana}, {Cardoso}, {Casassus}, {Catalano}, {Chamballu},
  {Chen}, {Chiang}, {Chiang}, {Christensen}, {Clements}, {Colombi}, {Colombo},
  {Couchot}, {Crill}, {Cuttaia}, {Danese}, {Davies}, {Davis}, {de Bernardis},
  {de Rosa}, {de Zotti}, {Delabrouille}, {D{\'e}sert}, {Dickinson}, {Diego},
  {Donzelli}, {Dor{\'e}}, {Dupac}, {En{\ss}lin}, {Eriksen}, {Finelli}, {Forni},
  {Franceschi}, {Galeotta}, {Ganga}, {G{\'e}nova-Santos}, {Ghosh}, {Giard},
  {Gonz{\'a}lez-Nuevo}, {G{\'o}rski}, {Gregorio}, {Gruppuso}, {Hansen},
  {Harrison}, {Helou}, {Hern{\'a}ndez-Monteagudo}, {Hildebrandt}, {Hivon},
  {Hornstrup}, {Jaffe}, {Jaffe}, {Jones}, {Keih{\"a}nen}, {Keskitalo},
  {Kneissl}, {Knoche}, {Kunz}, {Kurki-Suonio}, {L{\"a}hteenm{\"a}ki},
  {Lamarre}, {Lasenby}, {Lawrence}, {Leonardi}, {Liguori}, {Lilje},
  {Linden-V{\o}rnle}, {L{\'o}pez-Caniego}, {Mac{\'{\i}}as-P{\'e}rez}, {Maffei},
  {Maino}, {Mandolesi}, {Marshall}, {Martin}, {Mart{\'{\i}}nez-Gonz{\'a}lez},
  {Masi}, {Massardi}, {Matarrese}, {Mazzotta}, {Meinhold}, {Melchiorri},
  {Mendes}, {Mennella}, {Migliaccio}, {Miville-Desch{\^e}nes}, {Moneti},
  {Montier}, {Morgante}, {Mortlock}, {Munshi}, {Naselsky}, {Nati}, {Natoli},
  {N{\o}rgaard-Nielsen}, {Noviello}, {Novikov}, {Novikov}, {Oxborrow},
  {Pagano}, {Pajot}, {Paladini}, {Paoletti}, {Patanchon}, {Pearson}, {Peel},
  {Perdereau}, {Perrotta}, {Piacentini}, {Piat}, {Pierpaoli}, {Pietrobon},
  {Plaszczynski}, {Pointecouteau}, {Polenta}, {Ponthieu}, {Popa}, {Pratt},
  {Prunet}, {Puget}, {Rachen}, {Rebolo}, {Reich}, {Reinecke}, {Remazeilles},
  {Renault}, {Ricciardi}, {Riller}, {Ristorcelli}, {Rocha}, {Rosset},
  {Roudier}, {Rubi{\~n}o-Mart{\'{\i}}n}, {Rusholme}, {Sandri}, {Savini},
  {Scott}, {Spencer}, {Stolyarov}, {Sutton}, {Suur-Uski}, {Sygnet}, {Tauber},
  {Tavagnacco}, {Terenzi}, {Tibbs}, {Toffolatti}, {Tomasi}, {Tristram},
  {Tucci}, {Valenziano}, {Valiviita}, {Van Tent}, {Varis}, {Vielva}, {Villa},
  {Wandelt}, {Watson}, {Wilkinson}, {Ysard}, {Yvon}, {Zacchei}, \&
  {Zonca}}]{Planck2014_XV}
---. 2013, ArXiv e-prints, arXiv:1309.1357

\bibitem[{{Planck Collaboration} {et~al.}(2014{\natexlab{a}}){Planck
  Collaboration}, {Ade}, {Alves}, {Aniano}, {Armitage-Caplan}, {Arnaud},
  {Atrio-Barandela}, {Aumont}, {Baccigalupi}, {Banday}, {Barreiro}, {Battaner},
  {Benabed}, {Benoit-L{\'e}vy}, {Bernard}, {Bersanelli}, {Bielewicz}, {Bock},
  {Bond}, {Borrill}, {Bouchet}, {Boulanger}, {Burigana}, {Cardoso}, {Catalano},
  {Chamballu}, {Chiang}, {Colombo}, {Combet}, {Couchot}, {Coulais}, {Crill},
  {Curto}, {Cuttaia}, {Danese}, {Davies}, {Davis}, {de Bernardis}, {de Zotti},
  {Delabrouille}, {D{\'e}sert}, {Dickinson}, {Diego}, {Donzelli}, {Dor{\'e}},
  {Douspis}, {Dunkley}, {Dupac}, {En{\ss}lin}, {Eriksen}, {Falgarone},
  {Fanciullo}, {Finelli}, {Forni}, {Frailis}, {Fraisse}, {Franceschi},
  {Galeotta}, {Ganga}, {Ghosh}, {Giard}, {Gonz{\'a}lez-Nuevo}, {G{\'o}rski},
  {Gregorio}, {Gruppuso}, {Guillet}, {Hansen}, {Harrison}, {Helou},
  {Hern{\'a}ndez-Monteagudo}, {Hildebrandt}, {Hivon}, {Hobson}, {Holmes},
  {Hornstrup}, {Jaffe}, {Jaffe}, {Jones}, {Keih{\"a}nen}, {Keskitalo},
  {Kisner}, {Kneissl}, {Knoche}, {Kunz}, {Kurki-Suonio}, {Lagache}, {Lamarre},
  {Lasenby}, {Lawrence}, {Leahy}, {Leonardi}, {Levrier}, {Liguori}, {Lilje},
  {Linden-V{\o}rnle}, {L{\'o}pez-Caniego}, {Lubin}, {Mac{\'{\i}}as-P{\'e}rez},
  {Maffei}, {Magalh{\~a}es}, {Maino}, {Mandolesi}, {Maris}, {Marshall},
  {Martin}, {Mart{\'{\i}}nez-Gonz{\'a}lez}, {Masi}, {Matarrese}, {Mazzotta},
  {Melchiorri}, {Mendes}, {Mennella}, {Migliaccio}, {Miville-Desch{\^e}nes},
  {Moneti}, {Montier}, {Morgante}, {Mortlock}, {Munshi}, {Murphy}, {Naselsky},
  {Nati}, {Natoli}, {Netterfield}, {Noviello}, {Novikov}, {Novikov},
  {Oppermann}, {Oxborrow}, {Pagano}, {Pajot}, {Paoletti}, {Pasian},
  {Perdereau}, {Perotto}, {Perrotta}, {Piacentini}, {Pietrobon},
  {Plaszczynski}, {Pointecouteau}, {Polenta}, {Popa}, {Pratt}, {Rachen},
  {Reach}, {Reinecke}, {Remazeilles}, {Renault}, {Ricciardi}, {Riller},
  {Ristorcelli}, {Rocha}, {Rosset}, {Roudier}, {Rubi{\~n}o-Mart{\'{\i}}n},
  {Rusholme}, {Salerno}, {Sandri}, {Savini}, {Scott}, {Spencer}, {Stolyarov},
  {Stompor}, {Sudiwala}, {Sutton}, {Suur-Uski}, {Sygnet}, {Tauber}, {Terenzi},
  {Toffolatti}, {Tomasi}, {Tristram}, {Tucci}, {Valenziano}, {Valiviita}, {Van
  Tent}, {Vielva}, {Villa}, {Wandelt}, {Zacchei}, \& {Zonca}}]{Planck2014_XXII}
{Planck Collaboration}, {Ade}, P.~A.~R., {Alves}, M.~I.~R., {et~al.}
  2014{\natexlab{a}}, submitted to A\&A, arXiv:1405.0874

\bibitem[{{Planck Collaboration} {et~al.}(2014{\natexlab{b}}){Planck
  Collaboration}, {Ade}, {Aghanim}, {Alves}, {Arnaud}, {Ashdown},
  {Atrio-Barandela}, {Aumont}, {Baccigalupi}, {Banday}, {Barreiro}, {Battaner},
  {Benabed}, {Benoit-L{\'e}vy}, {Bernard}, {Bersanelli}, {Bielewicz}, {Bobin},
  {Bonaldi}, {Bond}, {Bouchet}, {Boulanger}, {Burigana}, {Cardoso}, {Catalano},
  {Chamballu}, {Chiang}, {Christensen}, {Clements}, {Colombi}, {Colombo},
  {Combet}, {Couchot}, {Crill}, {Cuttaia}, {Danese}, {Davies}, {Davis}, {de
  Bernardis}, {de Rosa}, {de Zotti}, {Delabrouille}, {Dickinson}, {Diego},
  {Donzelli}, {Dor{\'e}}, {Douspis}, {Dupac}, {Efstathiou}, {En{\ss}lin},
  {Eriksen}, {Finelli}, {Forni}, {Frailis}, {Franceschi}, {Galeotta}, {Ganga},
  {G{\'e}nova-Santos}, {Ghosh}, {Giard}, {Giardino}, {Giraud-H{\'e}raud},
  {Gonz{\'a}lez-Nuevo}, {G{\'o}rski}, {Gregorio}, {Gruppuso}, {Hansen},
  {Harrison}, {Henrot-Versill{\'e}}, {Herranz}, {Hildebrandt}, {Hivon},
  {Hobson}, {Hornstrup}, {Hovest}, {Huffenberger}, {Jaffe}, {Jaffe}, {Jones},
  {Keih{\"a}nen}, {Keskitalo}, {Kisner}, {Kneissl}, {Knoche}, {Kunz},
  {Kurki-Suonio}, {Lagache}, {L{\"a}hteenm{\"a}ki}, {Lamarre}, {Lasenby},
  {Lawrence}, {Leonardi}, {Liguori}, {Lilje}, {Linden-V{\o}rnle},
  {L{\'o}pez-Caniego}, {Lubin}, {Mac{\'{\i}}as-P{\'e}rez}, {Maino},
  {Mandolesi}, {Martin}, {Mart{\'{\i}}nez-Gonz{\'a}lez}, {Masi}, {Massardi},
  {Matarrese}, {Mazzotta}, {Meinhold}, {Melchiorri}, {Mendes}, {Mennella},
  {Migliaccio}, {Mitra}, {Miville-Desch{\^e}nes}, {Moneti}, {Montier},
  {Morgante}, {Mortlock}, {Munshi}, {Murphy}, {Naselsky}, {Nati}, {Natoli},
  {N{\o}rgaard-Nielsen}, {Noviello}, {Novikov}, {Novikov}, {Oxborrow},
  {Pagano}, {Pajot}, {Paladini}, {Paoletti}, {Pasian}, {Pearson}, {Peel},
  {Perdereau}, {Perrotta}, {Piacentini}, {Piat}, {Pierpaoli}, {Pietrobon},
  {Plaszczynski}, {Pointecouteau}, {Polenta}, {Ponthieu}, {Popa}, {Pratt},
  {Prunet}, {Puget}, {Rachen}, {Reach}, {Rebolo}, {Reich}, {Reinecke},
  {Remazeilles}, {Renault}, {Ricciardi}, {Riller}, {Ristorcelli}, {Rocha},
  {Rosset}, {Roudier}, {Rubi{\~n}o-Mart{\'{\i}}n}, {Rusholme}, {Sandri},
  {Savini}, {Scott}, {Spencer}, {Stolyarov}, {Strong}, {Sutton}, {Suur-Uski},
  {Sygnet}, {Tauber}, {Tavagnacco}, {Terenzi}, {Tibbs}, {Toffolatti}, {Tomasi},
  {Tristram}, {Tucci}, {Valenziano}, {Valiviita}, {Van Tent}, {Varis},
  {Vielva}, {Villa}, {Wade}, {Wandelt}, {Watson}, {Yvon}, {Zacchei}, \&
  {Zonca}}]{Planck2014_XXIII}
{Planck Collaboration}, {Ade}, P.~A.~R., {Aghanim}, N., {et~al.}
  2014{\natexlab{b}}, Submitted to A\&A, arXiv:1406.5093

\bibitem[{{Rafikov}(2006)}]{Rafikov2006}
{Rafikov}, R.~R. 2006, \apj, 646, 288

\bibitem[{{Rubi{\~n}o-Mart{\'{\i}}n} {et~al.}(2012){Rubi{\~n}o-Mart{\'{\i}}n},
  {L{\'o}pez-Caraballo}, {G{\'e}nova-Santos}, \& {Rebolo}}]{Rubino-Martin2012a}
{Rubi{\~n}o-Mart{\'{\i}}n}, J.~A., {L{\'o}pez-Caraballo}, C.~H.,
  {G{\'e}nova-Santos}, R., \& {Rebolo}, R. 2012, Advances in Astronomy, 2012,
  doi:10.1155/2012/351836

\bibitem[{{Scaife} {et~al.}(2008){Scaife}, {Hurley-Walker}, {Davies},
  {Duffett-Smith}, {Feroz}, {Grainge}, {Green}, {Hobson}, {Kaneko}, {Lasenby},
  {Pooley}, {Saunders}, {Scott}, {Titterington}, {Waldram}, \&
  {Zwart}}]{Scaife2008}
{Scaife}, A.~M.~M., {Hurley-Walker}, N., {Davies}, M.~L., {et~al.} 2008,
  \mnras, 385, 809

\bibitem[{{Scaife} {et~al.}(2009){Scaife}, {Hurley-Walker}, {Green}, {Davies},
  {Franzen}, {Grainge}, {Hobson}, {Lasenby}, {Pooley},
  {Rodr{\'{\i}}guez-Gonz{\'a}lvez}, {Saunders}, {Scott}, {Shimwell},
  {Titterington}, {Waldram}, \& {Zwart}}]{Scaife2009}
{Scaife}, A.~M.~M., {Hurley-Walker}, N., {Green}, D.~A., {et~al.} 2009, \mnras,
  400, 1394

\bibitem[{{Scaife} {et~al.}(2010{\natexlab{a}}){Scaife}, {Green}, {Pooley},
  {Davies}, {Franzen}, {Grainge}, {Hobson}, {Hurley-Walker}, {Lasenby},
  {Olamaie}, {Richer}, {Rodr{\'{\i}}guez-Gonz{\'a}lvez}, {Saunders}, {Scott},
  {Shimwell}, {Titterington}, {Waldram}, \& {Zwart}}]{Scaife2010b}
{Scaife}, A.~M.~M., {Green}, D.~A., {Pooley}, G.~G., {et~al.}
  2010{\natexlab{a}}, \mnras, 403, L46

\bibitem[{{Scaife} {et~al.}(2010{\natexlab{b}}){Scaife}, {Nikolic}, {Green},
  {Beck}, {Davies}, {Franzen}, {Grainge}, {Hobson}, {Hurley-Walker}, {Lasenby},
  {Olamaie}, {Pooley}, {Rodr{\'{\i}}guez-Gonz{\'a}lvez}, {Saunders}, {Scott},
  {Shimwell}, {Titterington}, {Waldram}, \& {Zwart}}]{Scaife2010a}
{Scaife}, A.~M.~M., {Nikolic}, B., {Green}, D.~A., {et~al.} 2010{\natexlab{b}},
  \mnras, 406, L45

\bibitem[{{Tibbs} {et~al.}(2013){Tibbs}, {Scaife}, {Dickinson}, {Paladini},
  {Davies}, {Davis}, {Grainge}, \& {Watson}}]{Tibbs2013}
{Tibbs}, C.~T., {Scaife}, A.~M.~M., {Dickinson}, C., {et~al.} 2013, \apj, 768,
  98

\bibitem[{{Tibbs} {et~al.}(2011){Tibbs}, {Flagey}, {Paladini}, {Compi{\`e}gne},
  {Shenoy}, {Carey}, {Noriega-Crespo}, {Dickinson}, {Ali-Ha{\"\i}moud},
  {Casassus}, {Cleary}, {Davies}, {Davis}, {Hirata}, \& {Watson}}]{Tibbs2011}
{Tibbs}, C.~T., {Flagey}, N., {Paladini}, R., {et~al.} 2011, \mnras, 418, 1889

\bibitem[{{Tibbs} {et~al.}(2012){Tibbs}, {Paladini}, {Compi{\`e}gne},
  {Dickinson}, {Alves}, {Flagey}, {Shenoy}, {Noriega-Crespo}, {Carey},
  {Casassus}, {Davies}, {Davis}, {Molinari}, {Elia}, {Pestalozzi}, \&
  {Schisano}}]{Tibbs2012}
{Tibbs}, C.~T., {Paladini}, R., {Compi{\`e}gne}, M., {et~al.} 2012, \apj, 754,
  94

\bibitem[{{Tielens}(2008)}]{Tielens2008}
{Tielens}, A.~G.~G.~M. 2008, \araa, 46, 289

\bibitem[{{Vidal} {et~al.}(2011){Vidal}, {Casassus}, {Dickinson}, {Witt},
  {Castellanos}, {Davies}, {Davis}, {Cabrera}, {Cleary}, {Allison}, {Bond},
  {Bronfman}, {Bustos}, {Jones}, {Paladini}, {Pearson}, {Readhead}, {Reeves},
  {Sievers}, \& {Taylor}}]{Vidal2011}
{Vidal}, M., {Casassus}, S., {Dickinson}, C., {et~al.} 2011, \mnras, 414, 2424

\bibitem[{{Watson} {et~al.}(2005){Watson}, {Rebolo},
  {Rubi{\~n}o-Mart{\'{\i}}n}, {Hildebrandt}, {Guti{\'e}rrez},
  {Fern{\'a}ndez-Cerezo}, {Hoyland}, \& {Battistelli}}]{Watson2005}
{Watson}, R.~A., {Rebolo}, R., {Rubi{\~n}o-Mart{\'{\i}}n}, J.~A., {et~al.}
  2005, \apjl, 624, L89

\bibitem[{{Ysard} {et~al.}(2011){Ysard}, {Juvela}, \& {Verstraete}}]{Ysard2011}
{Ysard}, N., {Juvela}, M., \& {Verstraete}, L. 2011, \aap, 535, A89

\bibitem[{{Ysard} \& {Verstraete}(2010)}]{Ysard2010a}
{Ysard}, N., \& {Verstraete}, L. 2010, \aap, 509, A12

\end{thebibliography}
\bibliographystyle{apj}

%\begin{thebibliography}{99}
%\end{thebibliography}

%%%% NOTES

\end{document}